\documentclass[12pt]{article}

\usepackage{amsmath}
\usepackage{amssymb}
\usepackage{amsfonts}
\usepackage{amscd}
\usepackage{amsbsy}
\usepackage{amsthm}
\usepackage{latexsym}
\usepackage{graphicx}
\usepackage{slashed}
\usepackage{color}

\DeclareMathOperator{\cx}{\square}

\renewcommand{\baselinestretch}{1.2}
\def\beq{\begin{eqnarray}}
\def\eeq{\end{eqnarray}}
\newcommand{\nn}{\nonumber}

\def\tr{\,\mbox{tr}\,}

\def\Tr{\,\mbox{Tr}\,}

\def\al{\alpha}
\def\be{\beta}

\def\ga{\gamma}
\def\de{\delta}
\def\vp{\varepsilon}
\def\ep{\epsilon}
\def\ze{\zeta}

\def\la{\lambda}
\def\na{\nabla}
\def\pa{\partial}

\def\si{\sigma}

\def\Ga{\Gamma}
\def\De{\Delta}

\def\Si{\Sigma}

\begin{document}

\begin{center}

{\large\bf
Antisymmetric Tensor Field and Cheshire Cat Smile
\\
of the Local Conformal Symmetry\footnote{Includes corrections in
formulas
compared to the published version [EPhJ C84 (2024) 108].   
}}
\vskip 5mm

\textbf{Ilya L. Shapiro}
\footnote{
E-mail address: \ ilyashapiro2003@ufjf.br}

\vskip 5mm

{\sl Departamento de F\'{\i}sica, ICE,
Universidade Federal de Juiz de Fora
\\
Campus Universit\'{a}rio - Juiz de Fora, 36036-900, MG, Brazil}
\end{center}

\vskip 6mm

\centerline{\textbf{Abstract}}
\vskip 1mm

\begin{quotation}
\noindent
The conformal version of the antisymmetric second-order tensor
field in four spacetime dimensions does not have gauge invariance
extensively discussed in the literature for more than half a century.
Our first observation is that, when coupled to fermions, only the
conformal version provides renormalizability of the theory at the
one-loop level. General considerations are supported by the derivation
of one-loop divergences in the fermionic sector, indicating good
chances for asymptotic freedom. The arguments concerning one-loop
renormalizability remain valid in the presence of self-interactions
and the masses for both fermion and antisymmetric tensor fields. In
the flat spacetime limit, even regardless the conformal symmetry has
gone, there is an expectation to meet renormalizability in all loop orders.
\vskip 3mm

\noindent
\textit{Keywords:} \ 
conformal symmetry, antisymmetric tensor field,
renormalization
\vskip 2mm

\noindent
\textit{MSC:} \ 
81T10,  
81T15,  
81T17,  
81T20   

\end{quotation}

\section{Introduction}
\label{SecIntro}

What we know about fundamental physics converges to the Minimal
Standard Model (MSM) with certain modifications related to the
neutrino masses. On the other hand, this knowledge comes from the
particle physics and the tests on accelerators and in the
high-precision experiments, both ways provide the information
restricted by available energy scales and possible large masses of
the particles (or bound states, or condensates) beyond MSM. The
possible particles beyond MSM may belong to the conventional
irreducible representations of the Lorentz group, i.e., scalars,
spinors and vectors. But there is also a possibility to have other
types of fields, such as an antisymmetric second order tensor, which
is the subject of the present work.

Starting from the 1966 paper \cite{OgiPolu67} (see also
\cite{KalbRamon}), the study of antisymmetric tensor fields
attracted a lot of interest, being an important part of the gauge
field theory. Let us start by mentioning a few relevant references.
Geometric formulation and relation to gravity have been discussed
in \cite{Damour1992}. Interaction of antisymmetric field to matter
and the formulation of the corresponding renormalizable theory was
given in \cite{Avdeev1993}. It is worth noting that the interaction
of a Dirac fermion with the electromagnetic field tensor $F_{\mu\nu}$
appears naturally in the effective context of QED (see, e.g.,
\cite{GitmanSaa} and references therein).

The antisymmetric tensor field represents an interesting theoretical
model in many respects (see, e.g.,
\cite{Pasti1995,Quevedo1996,Buch2008,Siegel1999}), including the
description of Lorentz violation \cite{Altschul2009,Albert}. In the
gravitational physics, antisymmetric fields may be a viable (albeit
not extensively explored yet) alternative to scalars and vectors in
the models of inflation \cite{Aashish2018} (see also most recent
work \cite{Panda2023} and further references therein). The last
aspect looks especially important owing to the existing
expectations to go far beyond the
framework of MSM using cosmological and astrophysical data
instead of laboratory experiments. In both (cosmology and laboratory)
cases, it is important to formulate the new field in a consistent way,
that should include interaction with matter (i.e., with fermions, in
the first place) and the possibility to incorporate quantum effects,
at least in the lowest nontrivial orders of the loop expansion.

In the quantum field theory framework, the effective action of
massive antisymmetric fields in curved spacetime was discussed, e.g.,
in \cite{Sezgin1980,Duff1980,Tiberio-antis,Aashish2018-int}. One
can say that many different issues related to antisymmetric fields
have been discussed and explored in the literature. In the present
work, we aim to explore a local conformal symmetry of the
second-order antisymmetric tensor field. Previously this issue
was addressed in relation to conformal supergravity
\cite{FrTs-superconf,Fradkin1985} where the conformal actions
similar (albeit not equivalent) to those calculated below were
obtained. Let us also mention the mathematical works (see, e.g.,
\cite{Branson,Erdmenger1997} and further references therein) and
the preprint \cite{Barbashov1983}, where the conformal operators
acting on $k$-forms were constructed and explored.
In what follows, we shall obtain the conformal action of this theory
with certain generalizations and also show how the requirement of
conformal invariance helps to establish the unique general form of
the action which is compatible with the quantum consistency. Anticipating
the result, let us say that this form does not possess the
gauge symmetry \cite{OgiPolu67} and, therefore, requires that the
antisymmetric tensor field propagates more than the minimal amount
of degrees of freedom required by the irreducible representation of
the Lorentz group.

In the following, we show how the conformal symmetry is operational
in formulating the consistent action of antisymmetric field coupled
to fermions, and confirm our arguments by deriving the one-loop
divergences coming from the quantum Dirac field loop. The rest of
the article is organized as follows.
In Sec.~\ref{sec2} we construct the expressions invariant under the
local conformal symmetry and discuss how this symmetry helps
to fix the general form of renormalizable action, including in the
case when this symmetry is broken by massive terms.
Sec.~\ref{sec3} reports on the derivation of one-loop divergences.
This calculation includes a number of reduction relations which
are collected in the Appendixes A and B. In Sec.~\ref{sec4} we
write down the action of renormalizable (at least at the one-loop
level) theory which follows from the considerations of the previous
sections and explore renormalization group equations in the new
renormalizable model. A short Sec.~\ref{Sec5} presents the
flat-spacetime limit of the theory and arguments of why this theory
may be all-loop renormalizable, even despite the guiding conformal
symmetry does not exist in the flat limit.
Finally, in Sect.~\ref{Conc} we draw our conclusions and describe
the possible extensions of the present work.

The analysis is restricted to the four-dimensional spacetime ($4D$).
The conventions include the signature $(+,-,-,-)$, regardless the
Wick rotation to the Euclidean space is assumed in the part of the
heat-kernel calculations. The definition of  the Riemann tensor is $\,R_{\,.\,\be\mu\nu}^{\al}=\Ga_{\,\be\nu,\,\mu}^{\al}
-\Ga_{\,\be\mu,\,\nu}^\al \,+\, ...$,
of the Ricci tensor
$R_{\al\be}=R_{\,.\,\al\mu\be}^\mu$, and the scalar
curvature $R=R_{\,\al}^\al$. Our notations for derivatives are
$\na A = A\na + (\na A)$, except those places where there cannot
be misunderstanding.

\section{Conformal theory of antisymmetric tensor field}
 \label{sec2}

Our first purpose is to construct the theory of antisymmetric tensor
field $B_{\mu\nu}=-B_{\nu\mu}$, possessing local conformal
symmetry in curved spacetime. The first step is to postulate the
transformation rule for this tensor. We define the conformal
transformations as
\beq
g_{\mu\nu} = \bar{g}_{\mu\nu}\,e^{2\si}\,,
\qquad
B_{\mu\nu} = \bar{B}_{\mu\nu}\,e^{\si}\,,
\qquad
\si = \si(x)\,.
\label{confBg}
\eeq
The indices of the tensors without bars are raised and lowered
using the metric, therefore
$B^{\mu\nu} = \bar{B}^{\mu\nu}\,e^{-3\si}$. We shall see, in
what follows,  that (\ref{confBg}) is a fortunate choice, providing
the global conformal symmetry for the bilinear action of the field
$B_{\mu\nu}$ with two derivatives. Let us, from the very beginning,
agree that the global transformation is when $\si$ is a constant,
while local transformation assumes $\si=\si(x)$. Indeed, local
conformal symmetry implies a global one, but not \textit{vice versa}.
However, one can separate the terms in the action which have a
global symmetry and then look for their combinations possessing
local symmetry.

In case the metric $g_{\mu\nu}$ and the antisymmetric tensor field
$B_{\mu\nu}$ are coupled to the set of matter fields $\Phi_i$ with
the conformal weights $k_i$, the condition of conformal symmetry
means that the conformal Noether identity should be satisfied,
\beq
   2 g_{\mu\nu}\,\frac{\de S_c}{\de g_{\mu\nu}}
\,+\,B_{\mu\nu}\,\frac{\de S_c}{\de B_{\mu\nu}}
+\, \sum_i k_i \Phi_i\,\frac{\de S_c}{\de \Phi_i}
\, = \, 0 \,,
\label{confNoether}
\eeq
where the transformation rules for the fields $\Phi_i$ are
\beq
\Phi_i \,=\,\bar{\Phi}_i  e^{k_i \si}\,.
\label{confPhi}
\eeq

Our purpose is to formulate the conformal action of the metric
and the field $B_{\mu\nu}$, and subsequently explore its
relevance for the one-loop and higher-loop renormalization.
The first step is to write down the
list of obvious local conformal invariants,  including
\beq
&&
W_1 = \sqrt{-g}  \,B^{\mu\nu}B^{\al\be}C_{\al\be\mu\nu},
\nn
\\
&&
W_2 = \sqrt{-g} \,(B_{\mu\nu}B^{\mu\nu})^2,
\nn
\\
&&
W_3 = \sqrt{-g} \, B_{\mu\nu}B^{\nu\al}B_{\al\be}B^{\be\mu},
\label{W123}
\eeq
and
\beq
W_{11} = \sqrt{-g} \, B^{\mu\al}B^{\nu\be}C_{\al\be\mu\nu}
= \frac12\,W_1.
\label{W11red}
\eeq
where $C_{\al\be\mu\nu}$ is Weyl tensor. The reduction formula
(\ref{W11red}) can be easily obtained using the cyclic identify for
this tensor, $C_{\al\be\mu\nu}+C_{\al\nu\be\mu}+C_{\al\mu\nu\be} = 0$.

The list of global conformal invariants includes
\beq
&&
K_1 = \sqrt{-g} \,B^{\mu\nu}B^{\al\be} R_{\mu\al} g_{\nu\be},
\nn
\\
&&
K_2 = \sqrt{-g} \, B_{\mu\nu}B^{\mu\nu}R,
\nn
\\
&&
K_3 = \sqrt{-g} \,(\na_\al B_{\mu\nu}) (\na^\al B^{\mu\nu})
 = \sqrt{-g} \,(\na_\al B_{\mu\nu})^2,
\nn
\\
&&
K_4 = \sqrt{-g} \,(\na_\mu B^{\mu\nu}) (\na^\al B_{\al\nu})
= \sqrt{-g} \,(\na_\mu B^{\mu\nu})^2 .
\label{K123}
\eeq
There are also three reducible expressions
\beq
&&
K_{11} = \sqrt{-g} \, B^{\mu\nu}B^{\al\be}R_{\mu\nu\al\be}
= 2K_1 - \frac13\,K_2 + W_1.
\nn
\\
&&
K_{12} = \sqrt{-g} \, B^{\mu\al}B^{\nu\be}R_{\mu\nu\al\be}
= \frac12\,K_{11}.
\nn
\\
&&
K_{31} = \sqrt{-g} \,(\na_\al B_{\mu\nu}) (\na^\mu B^{\al\nu})
= K_4 - \frac16\,K_2 + \frac12\,W_1
\nn
\\
&&
\qquad\qquad
\,+\,
\na_\al \big[ B_{\mu\nu} (\na^\mu B^{\al\nu})
- B^{\al\nu} (\na^\mu B_{\mu\nu}) \big].
\label{K31red}
\eeq

Consider the infinitesimal conformal variations of the irreducible
terms. The basic variations are (see, e.g., \cite{Stud})
\beq
&&
\de_c  \Ga^\la_{\al\be}
\,=\, \de^\la_{\,\al}\si_\be + \de^\la_{\,\be}\si_\al
- \bar{g}_{\al\be} \si^\la,
\nn
\\
&&
\de_c  R \,=\,  - 2\bar{R}\si - 6 \bar{\cx}\si,
\nn
\\
&&
\de_c  R_{\al\be}
\,=\,  - \bar{g}_{\al\be} \bar{\cx}\si - 2 \si_{\al\be},
\label{conbasic}
\eeq
where $\si_\al = \bar{\na}_\al \si$ and
$\si_{\al\be} = \bar{\na}_\al \bar{\na}_\be \si$.
The covariant derivatives with bars are constructed with
the corresponding metric $\bar{g}_{\al\be}$.

After certain algebra, we obtain the variations of the terms
(\ref{K123})
\beq
&&
\de_c K_1 = \sqrt{- \bar{g}}
\, \bar{B}^{\mu\nu}
\big[
   2 \si^\la (\bar{\na}_\la \bar{B}_{\mu\nu})
+ 2 \si_\nu (\bar{\na}^\la \bar{B}_{\mu\la})
+ 2 \si^\la (\bar{\na}_\nu \bar{B}_{\mu\la})
\big],
\nn
\\
&&
\de_c K_2 = \sqrt{-\bar{g}} \, \bar{B}^{\mu\nu}
\big[12  \si^\la (\bar{\na}_\la \bar{B}_{\mu\nu})\big],
\nn
\\
&&
\de_c K_3 = \sqrt{-\bar{g}} \,\bar{B}^{\mu\nu}
\big[
4 \si_\nu (\bar{\na}^\la \bar{B}_{\mu\la})
- 4 \si^\la (\bar{\na}_\nu \bar{B}_{\mu\la})
- 2 \si^\la (\bar{\na}_\la \bar{B}_{\mu\nu})
\big],
\nn
\\
&&
\de_c K_4 = \sqrt{-\bar{g}} \,\bar{B}^{\mu\nu}
\,\big[2 \si_\nu (\bar{\na}^\la \bar{B}_{\mu\la})\big].
\label{deltasK123}
\eeq

It is easy to see that there are only three different variations,
which can be called $Z_1$, $Z_2$ and $Z_3$. Then
\beq
&&
\de_c K_1 = 2Z_1 + 2Z_2 + 2Z_3,
\nn
\\
&&
\de_c K_2 = 12Z_1,
\nn
\\
&&
\de_c K_3 = - 2Z_1 + 4Z_2 - 4Z_3,
\nn
\\
&&
\de_c K_4 = 2Z_2,
\label{deltasK}
\eeq
from what follows the combination
$K_3 - 4K_4 + 2K_1 -\frac16\,K_2$ is invariant. Thus, we
arrive at the fourth (and the last) conformal invariant
\beq
&&
W_4 \,=\,
\sqrt{-g}\Big\{
(\na_\al B_{\mu\nu}) (\na^\al B^{\mu\nu})
- 4 (\na_\mu B^{\mu\nu}) (\na^\al B_{\al\nu})
\nn
\\
&&
\qquad
\quad
+ \,2 B^{\mu\nu}R_\nu^{\,\al}B_{\mu\al}
-\frac16 \,RB_{\mu\nu}B^{\mu\nu}
\Big\}.
\label{W4}
\eeq
The last expression is interesting in various respects, so we
can make a list.

\textit{i)} \ A linear combination of the terms (\ref{W123}) and
(\ref{W4}) with arbitrary coefficients represents a new conformal
theory in $4D$. The expression very similar to (\ref{W4}) was
obtained in \cite{FrTs-superconf} (see also \cite{Fradkin1985}).
Other previously known examples include two- and
four-derivative scalars formulated in \cite{Penrose64,Chernikov68}
and \cite{FrTs-superconf,Paneitz}, one-derivative and
three-derivative spinors and gauge vector field. It is also worth
mentioning the recent work \cite{Paci2023} on the conformal
theory of mixed-symmetry tensors.

One of the unusual features of the conformal model (\ref{W4})
is the presence of nonminimal interaction of the $CBB$-type
with an arbitrary coefficient. In all previously known examples
in $4D$, the requirement of local conformal symmetry did not
leave space for arbitrary nonminimal parameters. Such freedom
emerges only in  $6D$ for the six-derivative conformal operator
acting on a scalar \cite{Hamada,Int-6d}.

\textit{ii)}  \ Out of the known examples of conformal theories
mentioned in the first point, only the vector model possesses also
gauge symmetry. One could expect the same for the antisymmetric
field $B_{\mu\nu}$, but this is not the case.
It is known starting from the seminal paper by Ogievetsky and
Polubarinov \cite{OgiPolu67}, that the ``usual'' antisymmetric field
theory, in flat spacetime, has a gauge symmetry under the
transformation
\beq
\de b_{\mu\nu}\,=\, \pa_\mu \xi_\nu - \pa_\nu \xi_\mu ,
\label{gauge}
\eeq
where the vector field $\xi_\mu$ satisfies the condition
$\pa_\mu \xi^\mu = 0$.

The calculations of the divergences of the
vacuum effective action \cite{Duff1980,Tiberio-antis,Aashish2018-int}
had to deal with the degeneracy related to this symmetry, making
quantization and the calculations themselves being highly nontrivial
and interesting. Our third observation is that the conformal term
(\ref{W4}) differs from the gauge invariant theory and, in
particular, it is not degenerate.

Let us elaborate on this feature in some detail. One can see that
\beq
F_{\mu\nu\la}
\,=\, \na_\mu b_{\nu\la}
+ \na_\la b_{\mu\nu} + \na_\nu b_{\la\mu}
\,=\, \pa_\mu b_{\nu\la}
+ \pa_\la b_{\mu\nu} + \pa_\nu b_{\la\mu}.
\label{F3}
\eeq
The same replacement of partial derivative by the covariant one
can be performed in (\ref{gauge}) and the invariance of
$F_{\mu\nu\la}$ obviously holds. A small algebra shows that
the gauge invariant combination has the form
\beq
&&
\mathcal{L}_{inv}
\,=\,
\frac13\,\sqrt{-g} \,F_{\mu\nu\la}F^{\mu\nu\la}
\,=\,
\sqrt{-g} \big(K_3 - 2K_{31}\big)
\,=\,
\sqrt{-g} \Big(K_3 - 2K_4 + \frac13\,K_2 - W_1\Big)
\mbox{\qquad }
\nn
\\
&&
\qquad
\quad
= \,\,\sqrt{-g}\Big\{
(\na_\al b_{\mu\nu})^2   
- 2 (\na_\mu b^{\mu\nu})^2   
+ \frac13\, b_{\mu\nu}^2 R
- b_{\mu\nu}b_{\al\be}C^{\al\be\mu\nu}\Big\},
\label{FF}
\eeq
where we used the reduction formula (\ref{K31red}). Using the
approach utilized in the search for conformal symmetry, one can
show that the last Lagrangian is a unique gauge invariant
combination with at most two derivatives of the field.

It is easy to note that the linear combinations in (\ref{FF}) and
(\ref{W4}) are different. One can also verify that the bilinear form
of the gauge invariant Lagrangian  (\ref{FF}) is degenerate, while
the bilinear form of  the conformal invariant Lagrangian (\ref{W4})
is not degenerate. On the other hand, the conformal transformation
in the Lagrangian (\ref{FF}) gives
\beq
&&
\de_c\,\big(\sqrt{-g} F_{\mu\nu\la}F^{\mu\nu\la}\big)
\,=\,
6 \sqrt{-g} B_{\mu\nu} \si^\la
\big(
\na_\mu B_{\nu\la}
+ \na_\la B_{\mu\nu}
+ \na_\nu B_{\la\mu}
\big)\,\neq\,0,
\label{FF-conf}
\eeq
that shows its conformal non-invariance, opposite to (\ref{W4}).
Let us note that this difference can be seen already in the conformal
equations of motion derived in \cite{Barbashov1983}.

\textit{iii)}
\ It is known that the gauge invariant theory of the antisymmetric
 field is equivalent to the theory of a real scalar field and the
 classical \cite{OgiPolu67} and quantum \cite{Schwarz1978}
 levels (the last issue was extensively discussed in different
 frameworks, e.g., in \cite{Duff1980}, \cite{Grisaru1984} and
 \cite{Buchbinder1988}. On the other hand, since the conformal
 model (\ref{W4}) is different from (\ref{FF}), we should not
 expect the equivalence with the scalar theory. Thus,  the conformal
model (\ref{W4}) may be an interesting object of study at both
classical and quantum levels.

In cosmology, the new conformal theory may be interesting for
describing dark radiation or a basis for new models of inflation
or of the dark sectors of the matter contents of the late Universe.
One of the potentially useful features is that the interaction terms
in (\ref{W123}) open the possibility to have broken symmetry at
low (IR) or high (UV) energies, such that the same model may
have very different conformal properties in the UV and in the IR.
Another potentially interesting aspect is that the second order in
derivatives nonminimal term $W_1$ includes the Weyl tensor and
therefore is supposed to affect only the cosmic perturbations while
it decouples from the conformal factor of the metric.

In quantum theory one can use the new conformal model given
by an \textit{arbitrary} linear combination of the terms (\ref{W123})
and (\ref{W4}) to test the important universal features of the trace
anomaly, especially the uniformity of signs of the Weyl-squared
and Gauss-Bonnet terms (see, e.g., \cite{Duff-94,PoImpo,OUP}).
We hope to explore this part in the future works.

\textit{iv)}
 \ Since there are two alternative theories for the antisymmetric
tensor field $B_{\mu\nu}$, one can ask which of these theories,
i.e. (\ref{FF}) or  (\ref{W4}), is ``better''. Of course, there is no
unique answer to this question, as it depends on the criteria of
the choice. However, as we shall see, there are important
aspects in which the model  (\ref{W4}), based on the conformal
symmetry has an advantage.

If considering a conformal theory of another field (e.g., a massless
fermion) on the background of an arbitrary metric and antisymmetric
field $B_{\mu\nu}$, we should expect the conformal invariance of the
one-loop divergences (more precise, the $4D$ limit of the coefficient
of the pole in dimensional regularization), as it was proved in
\cite{tmf}. In the present case, this means that the logarithmic
one-loop divergences  represent a linear combination of the
conformal terms. These terms include the square of the Weyl tensor,
Gauss-Bonnet topological term, total derivative term $\cx R$, the
conformal terms (\ref{W123}) and (\ref{W4}) constructed from
$B_{\mu\nu}$ and the metric, plus the total derivatives constructed
from the same fields.

One can couple $B_{\mu\nu}$ to the Dirac fermion as follows:
\beq
S_{1/2}
\,=\, i\int d^4x\sqrt{-g}\,\bar{\psi}
\big\{\ga^\mu\na_\mu  - \Sigma^{\mu\nu}B_{\mu\nu} - im
\big\} \psi,
\label{action-Dirac-B}
\eeq
where gamma-matrices are defined in a usual way as
$\ga^\mu = e^\mu_{\,a}\ga^a$, \ $\Si^{\mu\nu}
= \frac{i}{2}(\ga^\mu \ga^\nu - \ga^\nu \ga^\mu)$ and $m$ is
the mass of the spinor field. The massless version of this theory
possesses conformal symmetry under (\ref{confBg}) plus the
standard transformations for the fermions in $4D$,
\beq
\psi = \psi_*\,e^{- \frac32 \,\si}\,,
\qquad
\bar{\psi} = \bar{\psi}_*\,e^{- \frac32 \,\si}\,.
\label{conf_ferm}
\eeq
Making the comparison with (\ref{confPhi}), the last formula
implies that the conformal weight of the fermions $\psi$ and
$\bar{\psi}$ \  is \ $k_f = - 3/2$.

According to \cite{tmf}, the conformal symmetry should hold in the
one-loop counterterms. Therefore, in the massless case, the one-loop
divergences should be of the form (\ref{W123}) and (\ref{W4}), plus
surface terms, and not of the form (\ref{FF}). On top of that, the
presence of mass means that the violation of the conformal symmetry
is soft. As a result, even in the massive case, when $m \neq 0$, the
mass-independent one-loop divergences are expected to be exactly
as in the massless theory, i.e., a linear combination of (\ref{W123})
and (\ref{W4}).

An additional detail is that fermionic action (\ref{action-Dirac-B})
does not possess symmetry with respect to the gauge transformation
(\ref{gauge}). Instead, the mass-independent part of this action is
invariant under the local conformal transformations (\ref{confBg})
and (\ref{conf_ferm}). This means the conformal symmetry plays
a critical (guiding) role in the construction of renormalizable
theory of the antisymmetric field coupled to fermions. Therefore,
applications of these fields that do not rule out the interaction of
$B_{\mu\nu}$ with leptons and quarks, should be based on the
Lagrangian given by a linear combination of (\ref{W123}),
(\ref{W4}) and the mass-dependent terms, instead of the gauge
invariant expression (\ref{FF}). In the next section, we shall verify
this statement by deriving the one-loop divergences in the theory
(\ref{action-Dirac-B}).

\section{One-loop divergences for fermion fields}
 \label{sec3}

In order to check the relation between conformal symmetry and the
one-loop renormalizability, let us derive the one-loop divergences
for the Dirac fermion coupled to external metric and antisymmetric
field ${B}_{\mu\nu}$. The starting point will be the action
(\ref{action-Dirac-B}). Thus, our purpose is to evaluate the
divergent part of  of the expression
\beq
\bar{\Ga}(g,B)
\,=\, - \,i \Tr \log \hat{H},
\label{barGa}
\eeq
where
\beq
 \hat{H}  \,=\,
 \ga^\mu\na_\mu - \Si_{\mu\nu} B_{\mu\nu} + im\,.
\label{H}
\eeq
To reduce (\ref{barGa}) to the standard form, we can introduce
the conjugate operator. The simplest choice is to change only the
sign of the mass term,
\beq
 \hat{H}^*  \,=\,
 \ga^\mu\na_\mu - \Si_{\mu\nu} B_{\mu\nu} - im
\label{Hstar}
\eeq
and take into account that $\Tr \log \hat{H} = \Tr \log \hat{H}^*$
(see, e.g., \cite{OUP}). After certain algebra (see Appendix A for
necessary details concerning gamma-matrices), we get
\beq
\bar{\Ga}(g,B)
\,=\, - \,\frac{i}{2}\, \Tr \log \hat{F},
\label{GaF}
\eeq
where
\beq
\hat{F} \,=\, \hat{H} \hat{H}^*
\,=\,
\hat{1}\cx + 2\hat{h}^\al \na_\al + \hat{\Pi}.
\label{F}
\eeq
Starting from this point, we sometimes omit the symbol of
the unit matrix $\hat{1}$ and hats over the operators. The
remaining elements of the operator can be reduced to the forms
\beq
&&
\hat{h}^\al \,=\, 2 \ga^5 \ga_\be \tilde{B}^{\al\be},
\nn
\\
&&
\hat{\Pi}\,=\, m^2 - \frac14\,R + 2B_{\al\be} B^{\al\be}
- 2i (\na_\al B^{\al\be})\ga_\be
- \,2i \ga^5 B_{\al\be} \tilde{B}^{\al\be}
\nn
\\
&&
\qquad \qquad
+ \,\, 2 \ga^5  (\na_\al \tilde{B}^{\al\be})\ga_\be
- 4 i B_{\al\be}B_{\mu\nu} \Si^{\mu\al}g^{\nu\be},
\label{hPi}
\eeq
where the last term vanishes and we use the standard notation
for the dual tensor,
\beq
\tilde{B}_{\mu\nu}
\,=\,\frac12\,\vp_{\mu\nu\al\be} B^{\al\be}\,.
\label{dualB}
\eeq

The one-loop divergences are given by the standard heat-kernel
expression \cite{DeWitt65} (see also \cite{OUP} for introduction
and further references), with the sign corresponding to the
odd Grassmann parity of the quantum field,
\beq
{\bar \Ga}^{(1)}_{div}
&=&
\frac{\mu^{n-4}}{\vp}\,
\int d^nx\sqrt{-g}\,\tr
\Big\{ \frac{{\hat 1}}{180}\,(R_{\mu\nu\al\be}^2
- R_{\al\be}^2 + {\Box}R)
\nn
\\
&&
\qquad 
+ \,\,\frac12 {\hat P}^2
\,+\, \frac{1}{12}{\hat S}_{\mu\nu}^2
\,+\,\frac16 ({\Box}{\hat P})
\Big\},
\label{EAdivs}
\eeq
where $\vp = (4\pi)^2(n-4)$ is the parameter of dimensional
regularization and the operators ${\hat P}$ and
${\hat S}_{\mu\nu}$ are defined as
\beq
&&
\hat{P}\,=\, \hat{\Pi} +  \frac{{\hat 1}}{6}\, R
- \na_\mu{\hat h}^\mu - {\hat h}_\mu{\hat h}^\mu ,
\label{Pgen}
\nn
\\
&&
{\hat S}_{\mu\nu}\,=\,\hat{\mathcal R}_{\mu\nu}
+ \na_\nu{\hat h}_\mu-\na_\mu{\hat h}_\nu
+ {\hat h}_\nu{\hat h}_\mu-{\hat h}_\mu{\hat h}_\nu.
\label{Smngen}
\eeq
Here $\hat{\mathcal R}_{\mu\nu} = [\na_\nu ,\na_\mu]$ in
the corresponding space. In the present case, it is the space of
Dirac spinors and therefore we get the expressions
\beq
&&
\hat{P}
\,=\,
m^2 - \frac{1}{12}\,R
- 2 B^2_{\mu\nu}
- 2i (\na_\mu B^{\mu\nu})\ga_\nu
- 2i B_{\mu\nu}\tilde{B}^{\mu\nu}\ga^5
\label{P}
\label{Smn}
\\
&&
{\hat S}_{\mu\nu}
\,=\,
- \,\frac14\, R_{\mu\nu\rho\si}\ga^\rho \ga^\si
+ 2 \ga^5 \ga^\la \big(\na_\nu \tilde{B}_{\mu\la}
- \na_\mu \tilde{B}_{\nu\la}\big)
+ 4 \big(\ga^\la\ga^\tau - \ga^\tau\ga^\la \big)
\tilde{B}_{\mu\la} \tilde{B}_{\nu\tau}\,.
\nn
\eeq

Using these building blocks in (\ref{EAdivs}) is simple after
deriving the set of reduction formulas presented in Appendix B.
The result is the expression for the one-loop divergences
\beq
{\bar \Ga}^{(1)}_{div}
&=&
-\,\frac{\mu^{n-4}}{\vp}\int d^nx\sqrt{-g}\,
\Big\{
\frac43\big(W_1 - W_4 \big)
- \frac{8}{3}\,W_2
+ \frac{32}{3}\,W_3
+ 8m^2 B^2_{\mu\nu}
\nn
\\
&&
- \,\,2m^4
+ \frac13\,m^2 R
+ \frac{1}{20}\,C_{\mu\nu\al\be}^2
- \frac{11}{360}\,E_4
+ \frac{1}{30}\,{\Box}R
+ \frac{4}{3}\,{\Box} B^2_{\mu\nu}
\Big\},
\label{EA-Bg-divs}
\eeq
where we use the condensed notations (\ref{W123}), (\ref{W4})
and denote the Gauss-Bonnet integrand (Euler density in $4D$) as
$E_4 = R_{\mu\nu\al\be}^2 - 4R_{\mu\nu}^2 + R^2$.

Without the $B_{\mu\nu}$-dependent terms, the expression for the
divergences has the standard well-known form (see, e.g., \cite{birdav}).
When the field $B_{\mu\nu}$ is present, we note that there are all
the terms expected from the dimension and covariance arguments.
Remarkably, four of these terms are in the conformal combination
(\ref{W4}). It is certainly instructive that the mass-free
$B_{\mu\nu}$-dependent terms form conformal invariants $W_{1,2,3,4}$.
This property is owing to several cancelations, that can be explained
only by the effect of conformal symmetry. The result confirms
aforementioned theorem \cite{tmf} concerning the conformal
invariance of the one-loop divergences in the classically conformal
invariant theory. Only the mass-dependent terms in the $4D$ limit
of the integral in (\ref{EA-Bg-divs}) violate Noether identity
(\ref{confNoether}).

\section{Renormalization group equations}
 \label{sec4}

Taking into account the considerations presented above, we can
formulate the curved-space action of the antisymmetric field
$B_{\mu\nu}$ which produced a renormalizable theory at the
one-loop level. Such an action has the form
\beq
&&
S_B \,=\, \int d^4x \sqrt{-g}\,\Big\{
\frac12\big(W_4 + \la W_1\big)
- \frac12M^2 B_{\mu\nu}
- \frac{1}{4!}\big(f_2 W_2  + f_3 W_3
+ \ze \cx B_{\mu\nu}
\big)
\nn
\\
&&
\qquad \qquad \qquad
\,+\, \mbox{total derivatives} \Big\}
\,+\,S_g.
\label{actionB}
\eeq
Here $\la$ is an arbitrary nonminimal parameter of interaction
with the Weyl tensor and $f_{2,3}$ are arbitrary parameters of
the quartic self-coupling of the antisymmetric field. We included
one of the possible total derivatives for the sake of generality,
but renormalizability in higher loops may require more such
terms, especially the ones quoted in Eq.~(\ref{K31red}). The
coefficient in front of $W_4$ is chosen positive to avoid the
ghost states with negative kinetic energy. In this paper, we do
not discuss in full details the surface terms related to total
derivatives in the action (\ref{actionB}). These terms will be
considered in the subsequent work devoted to the conformal
(trace) anomaly and to the corresponding ambiguities. Finally, the
term $S_g$ is the usual metric-dependent vacuum action, which
was discussed in many ways (see, e.g., \cite{birdav} and
\cite{OUP}), so we avoid repeating well-known things here.

Consider the renormalization and renormalization
group running at the one-loop level. For this, let is rewrite the
action of fermions (\ref{action-Dirac-B}), introducing new
coupling constant $g$,
\beq
S_{1/2}
\,=\,
i\int d^4x\sqrt{-g}\,\bar{\psi}
\big\{
\ga^\mu\na_\mu  - g \Sigma^{\mu\nu}B_{\mu\nu} - im
\big\}
\psi.
\label{action-Dirac-Blam}
\eeq
Then, omitting purely metric-dependent terms, the expression for
the  $B_{\mu\nu}$-dependent part of the divergences
(\ref{EA-Bg-divs}) becomes
\beq
{\bar \Ga}^{(1)}_{div}(B,g)
&=&
-\,\frac{\mu^{n-4}}{\vp}\int d^nx\sqrt{-g}\,
\Big\{
\frac12\big(k_1W_1 + k_4W_4 \big)
\nn
\\
&&
\quad
- \,\,\frac{1}{4!}\big(k_2 W_2  + k_3 W_3 \big)
- \frac12 k_M B^2_{\mu\nu}
+ k_\ze \cx B_{\mu\nu}
\Big\},
\label{EA-Bg-divslam}
\eeq
where
\beq
&&
k_1 \,=\,k_4 \,=\, - \frac{8g^2}{3},
\quad\,\,
k_2 \,=\,64 g^4,
\quad\,\,
k_3 \,=\, -\,256 g^4,
\nn \\ &&
k_M \,=\,-\,16 g^2 m^2,
\qquad\,\,
k_\ze \,=\,\frac{4 g^2}{3}\,.
\label{ks}
\eeq

To elucidate the construction presented above,
let us consider the renormalization group equations that correspond
to the coefficients (\ref{ks}). As usual, we require the equality
of the bare and renormalized actions $S_0 = S + \De S = S_R$, where
the local counterterms are $\De S = - {\bar \Ga}^{(1)}_{div}(B,g)$.
All subsequent formulas are restricted to the one-loop approximation,
which means, in particular, $\mathcal{O}(1/\vp)$, while the higher
orders are neglected.

From the term $W_4$, we get the renormalization relation between
bare $B^{(0)}_{\mu\nu}$ and renormalized $B_{\mu\nu}$ fields,
\beq
B^{(0)}_{\mu\nu}\,=\,\mu^{\frac{n-4}{2}}
\Big(1 + \frac{k_4}{2\vp} \Big) B_{\mu\nu}.
\label{Ren_B}
\eeq
At this point, we meet an important difference with other theories,
such as QED. For the coupling constant $g$ we require
\beq
g_0B^{0}_{\mu\nu}\,=\,g B_{\mu\nu}.
\label{Ren_main}
\eeq
The mentioned difference is that, in the present case, there is no
gauge symmetry that protects the product $g B_{\mu\nu}$ and hence
this relation is based only on the possibility of hiding $g$ inside the
field and attribute the running to the logarithmic form factor for
the $W_4$ term. Let us note that the corresponding symmetry may
be eventually formulated, as it will be discussed in the
Conclusions. As a result of (\ref{Ren_main}), in the one-loop
approximation we get
\beq
g^0\,=\,\mu^{\frac{n-4}{2}}\Big(1 - \frac{k_4}{2\vp} \Big) g.
\label{Ren_g}
\eeq
Starting from this point, the considerations are quite standard.
After small algebra, we arrive at the renormalization relations
for the parameters $\la$, $f_2$ and $f_3$ in the form
\beq
&&
\la^{0}\,=\,\la + \frac{k_1 - k_4\la}{\vp}\,,
\nn
\\
&&
f_{2,3}^{0}\,=\,\mu^{4-n}
\Big(f_{2,3} \,+\, \frac{k_{2,3} - 2 k_4 f_{2,3}}{\vp} \Big),
\label{Ren_f2f3}
\eeq
and for the massive and surface terms
\beq
&&
M^2_0\,=\,M^ 2 + \frac{k_M - k_4M^2}{\vp} \,,
\nn
\\
&&
\ze_{0}\,=\,\ze + \frac{k_\ze - k_4\ze}{\vp}\,.
\label{Ren_Mze}
\eeq

The beta functions are defined as usual, e.g.,
\beq
\be_g \,=\, \lim_{n \to 4} \mu\frac{dg}{d\mu}.
\label{beta_g}
\eeq
Direct calculations give
\beq
\mu\frac{dg}{d\mu}  \,=\,  \frac{n-4}{2}g
\,+\, g\, \frac{dk_4}{dg}\,=\,
\frac{n-4}{2}g \,-\, \frac{4}{3(4\pi)^2}\,g^3.
\label{beta_gn}
\eeq
Let us give the list of similar expressions for the parameter
$\la$ and coupling constants,
\beq
&&
\mu\frac{d\la}{d\mu}
\,=\,
\frac{1}{2(4\pi)^2}
\Big( g \,\frac{d k_1}{d g} \,-\, \la g \,\frac{d k_4}{d g}\Big)
\,=\, \frac{8}{3(4\pi)^2}\,g^2 (\la - 1)\,,
\nn
\\
&&
\mu\frac{d \,f_{2,3}}{d\mu}
\,=\,  (n-4)f_{2,3} \,+\,  \frac{1}{(4\pi)^2}
\Big( k_{2,3} \,+\, f_{2,3}\, g \,\frac{d k_4}{d g} \Big).
\label{beta_laf2f3}
\eeq
The derivation of the relations (\ref{beta_gn}) and
(\ref{beta_laf2f3}) does not depend on the specific form of the
coefficients $k_{1,2,3,4}$ and can be useful in the theory with
quantum $B_{\mu\nu}$.

Consider the solutions of the renormalization group
equations. For compactness, we denote $t=\log (\mu/\mu_0)$
and attribute index zero to the values of running quantities at the
fiducial scale $\mu_0$. For the square of the running coupling
$g(t)$ we meet
\beq
\frac{dg^2}{dt}  \,=\, - \, \frac{8}{3(4\pi)^2}\,g^4
\,=\, - \, a^2\,g^4,
\qquad
a^2 = \frac{1}{6\pi^2}.
\label{RG_g}
\eeq
This is the typical equation for the asymptotic freedom in
the UV, with the solution
\beq
g^2(t)  \,=\, \frac{g_0^2}{1 + a^2 g_0^2 \,t}\,.
\label{RG_g_sol}
\eeq
The following observation is in order. The
self-interaction terms of the field $B_{\mu\nu}$ have fourth
powers of this field. This means the following consequence.
According to the power counting, even if we add the contributions
of the quantum antisymmetric tensor field, at the one loop level
the equation (\ref{RG_g}) and the solution (\ref{RG_g_sol}) do
not modify. Of course, this concerns only the effective charge
$g(\mu)$. The renormalization group equations for other
couplings, namely $\la$, $f_1$ and $f_2$ will get modified.

For the running kinetic nonminimal parameter $\la(t)$ we get
\beq
\la(t)  \,=\, 1 + (\la_0 - 1)\big(1 + a^2 g_0^2 \,t\big)\,,
\label{RG_la_sol}
\eeq
that means $\la = 1$ is UV-unstable fixed point. For an arbitrary
initial value $\la_0 \neq 1$, this parameter logarithmically runs to
infinity with the sign defined by the one of $\la_0 - 1$.

For the running couplings $f_{2,3}(t)$ we need to solve the
equations
\beq
&&
\frac{d \,f_{2,3}}{dt}
\,=\,  C_{2,3}g^4 \,-\, 2a^2 g^2f_{2,3},
\qquad
C_{2,3}\,=\,\frac{1}{(4\pi)^2}(64,\,- 256).
\label{RG_f2f3}
\eeq
This equation can be explored using the standard trick
\cite{Cheng1973}. For the ratios between $f_{2,3}(t)$ and
$g^2(t)$ we get the solutions
\beq
\bar{f}_{2,3}(t)
\,=\, \frac{f_{2,3}(t)}{g^2(t)}
\,=\, \Big(\bar{f}^0_{2,3} - \frac{C_{2,3}}{a^2} \Big)
\big[1 + a^2 g_0^2 \,t\big]^{-1}\,,
\label{RG_f2f3_sol}
\eeq
indicating UV stable fixed points at the values $\bar{f}_{2,3}
= \frac{C_{2,3}}{a^2}$. Let us note that the fixed point for
$\bar{f}_3$ has a negative sign, which means possible problems
with the stability of effective potential in the UV. Indeed, this
feature may change after taking the quantum effects of the
self-interactions into account, as this is also possible for the
scalar potential \cite{Cheng1973}. In any case, within the given
approximation, both running couplings $f_{2,3}(t)$ manifest
asymptotic freedom behavior in the UV, independent of the
sign of the coefficients  $C_{2,3}$.

The beta functions for $M^2$ and $\ze$ have the form
\beq
&&
\mu\frac{dM^2}{d\mu}
\,=\,
\frac{1}{2(4\pi)^2}
\Big(M^2 g \,\frac{d k_4}{d g}
\,-\, g \,\frac{d k_M}{d g}\Big)
\,=\, -\,g^2a^2 \big(M^2 - 6m^2 \big)\,,
\nn
\\
&&
\mu\frac{d\ze}{d\mu}
\,=\,
\frac{1}{2(4\pi)^2}
\Big(\ze g\,\frac{d k_4}{d g}
\,-\,  g \,\frac{d k_\ze}{d g}\Big)
\,=\, - g^2a^2\,
\Big( \ze + \frac12 \Big)\,.
\label{beta_Mze}
\eeq
The solutions of the corresponding equations can be easily found
and are similar to (\ref{RG_f2f3_sol}), but we skip these formulas
here, as they are not very informative.

\section{Flat limit and renormalizability}
\label{Sec5}

In the flat limit, the sum of $N$ copies of the fermionic action
(\ref{action-Dirac-Blam}) and (\ref{actionB}), gives
\beq
&&
S_{flat} \,=\, \int d^4x\,\Big\{
\sum_{k=1}^N i \bar{\psi}_k
\big(\ga^\mu\na_\mu  - g \Sigma^{\mu\nu}B_{\mu\nu} - im
\big)\psi_k
\,+\, 
\frac12 (\pa_\al B_{\mu\nu})^2  
\nn
\\
&&
\qquad \qquad
- \,\, 2 (\pa_\mu B^{\mu\nu})^2  
- \frac12M^2 B^2_{\mu\nu}
- \frac{f_2}{4!}  (B^2_{\mu\nu})^2
- \frac{f_3}{4!}  B_{\mu\nu}B^{\nu\al}B_{\al\be}B^{\be\mu}
\Big\}\,,
\label{actionflat}
\eeq
where all indices are raised and lowered with the Minkowski
metric. The flat-spacetime version of our model gives the
possibility to consider several sides of the renormalizability
problem, so it is useful to elaborate a list of arguments.

\textit{i)} \ Renormalizability of the theory depends on the types
of the counterterms required to cancel divergences in the given
order of the loop expansion. The UV divergences in quantum field
theory are always removed by adding local counterterms and the
model (\ref{actionflat}) is not supposed to be an exception.

\textit{ii)} \ The possible violation of renormalizability in
conformal quantum theory on a curved spacetime is related to the
violation of the conformal symmetry because of the trace anomaly.
The general structure of anomaly, including of the effective action
induced by anomaly are pretty well-known (see, e.g., \cite{OUP}
for a review). The anomaly-induced action \cite{rie,FrTs84} includes
nonlocal and local terms, usually the last are considered irrelevant.
On the other hand, it is unlikely that the nonlocal terms produced
in the subdiagrams produce local divergent terms in the superficial
integration. Thus, the locality of the UV divergences may provide
the nonlocal terms in the effective action being non-important for
renormalizability. On top of this, we know the structure of the
nonlocal terms (see, e.g., \cite{Int-6d,OUP,PoImpo,Mottola-2017})
and these terms certainly vanish in the flat background. Therefore,
the relevant ones for renormalizability (certainly, in the flat limit)
are only the local terms.

\textit{iii)} \ The local terms in the anomaly-induced action usually
depend on the scheme of renormalization, as it was recently discussed
in \cite{anomaly-2004} for scalar fields and in \cite{AtA} for the
axial vector related to torsion.\footnote{This seems to be the closest
analog to the case of $B_{\mu\nu}$, especially in the part concerning
local terms.}
Assuming that the same is true for the theory of
antisymmetric field, there may be a renormalization scheme that
provides the absence of, at least curvature-independent, nonconformal
local terms at higher loops. In this case, the theory (\ref{actionflat})
may be all-loop renormalizable.

\textit{iv)} \  The amusing feature of the action (\ref{actionflat})
is that there is no local conformal symmetry in the flat limit. But,
on the other hand, the ``conformal'' restriction on the coefficients
of the two kinetic terms in (\ref{actionflat})  i.e.,
$(\pa_\al B_{\mu\nu})^2$ and $(\pa_\mu B^{\mu\nu})^2$,
still holds in the one-loop divergences and, according to the
previous point, may actually hold even beyond the one-loop
approximation, at least in an appropriate renormalization
scheme.\footnote{The elementary evaluation of
power counting shows that the renormalizability holds at the
two-loop level.}
This situation resembles the smile of Cheshire Cat, which remains
seen even when the Cat itself has gone.

\textit{v)} \  According to the general theory of interacting
fields in curved spacetime (see, e.g., \cite{tmf,PoImpo,OUP}), the
renormalization of the curvature-independent ``minimal'' terms is the
same in flat and curved spaces. Therefore, if the relation between
$(\pa_\al B_{\mu\nu})^2$ and $(\pa_\mu B^{\mu\nu})^2$ really holds
beyond one-loop order in flat spacetime, it will be the same for the
curved-space analogs $(\na_\al B_{\mu\nu})^2$ and
$(\na_\mu B^{\mu\nu})^2$. Then the loss of renormalizability
may occur only owing to the curvature-dependent terms, violating
the form of the conformal term (\ref{W4}).

\section{Conclusions and discussions}
\label{Conc}

We constructed the one-loop renormalizable theory of self-interacting
and interacting with fermions, antisymmetric tensor field. Different
from the Abelian or non-Abelian vector fields, there is no usual
gauge invariance in our model, such that the main symmetry is
local conformal invariance. The invariant action includes the terms
(\ref{W4}) with fixed coefficients, a qualitatively new nonminimal
term $W_1$ and the two self-interaction terms $W_{2,3}$, which
may be used for spontaneous or dynamical symmetry breaking.

The conformal
invariance requires the presence of the local metric of spacetime
but, even in a flat limit, the effect of the conformal symmetry is
sufficient to provide that the relation between two kinetic terms
in the action (\ref{actionflat}) holds at the quantum level. On top
of this, there are serious arguments in favor of all-loop
renormalizability of the theory, at least in the flat limit and using
a specially tuned scheme of renormalization. Our general
considerations are confirmed by the one-loop calculation of the
fermionic contributions to the divergent part of the effective action
of antisymmetric field and metric. These calculations indicate the
possibility of the asymptotic freedom in the theory, but this feature
should be verified by deriving the divergences in a full theory,
including quantum field $B_{\mu\nu}$. We expect to clarify
this issue elsewhere.

In this first paper, we leave unexplored several aspects of the new
theory, that is supposed to be done in future works. First of all, it
would be interesting to write down the trace anomaly and the
corresponding effective action. It is especially interesting to
consider the ambiguity in the local terms in the anomaly-induced
action since this may be relevant for better understanding
renormalizability beyond the one-loop order. Another obvious
problem to solve
is the derivation of full divergences, including the contributions
of the proper field $B_{\mu\nu}$, as mentioned above. After this
calculation, one can draw a more definite conclusion about the
UV limit in this theory, including the asymptotic freedom. At the
present stage, we can only claim the asymptotic freedom for a
sufficiently large amount of the fermion fields when the fermionic
contributions (\ref{RG_g}) should dominate.

One more interesting aspect is a possible link to the theory of
irreducible tensor field $b_{\mu\nu}$ and its action (\ref{FF}).
It is well-known that the free version of the gauge
invariant antisymmetric field is equivalent to the scalar theory
\cite{OgiPolu67}), but such an equivalence does not hold for the
conformal version, which is considered in the present work.
It might happen that there is a St$\ddot{\rm u}$ckelberg-like
procedure linking two kinds of fields. Elaborating on this issue
may be helpful, in particular, for a better understanding of the
relation (\ref{Ren_main}). Another potentially interesting aspect
of the problem is that, different from the theory (\ref{FF}),
the model (\ref{actionB}) may have problems with the
negative-energy states, i.e., ghosts, as discussed in
\cite{Avdeev1994}. In this case, we shall meet
a new example of the theory which is renormalizable and not
unitary, at least at the tree level. Since this theory looks simpler
than, e.g., higher derivative quantum gravity, it may serve as
a useful model to explore the issues such as instabilities in the
classical solutions and resolution of the problem of ghosts
at the quantum level.
This feature makes the theory of renormalizable
antisymmetric field worth investigating. Another possible
application is to consider this model in the effective approach,
where it has various interesting phenomenological applications
to particle physics and astrophysics
(see, e.g., \cite{Chizhov2017} and further references therein).

Last, but not least, it would be interesting to explore some
applications of the new model (\ref{actionB}) to cosmology and
maybe even to particle physics. Let us note that the model of
an antisymmetric field without gauge invariance has been already
applied to inflation in \cite{Aashish2018, Panda2023}. Our
analysis may be useful in fixing the coefficients of the terms
$(\na_\al B_{\mu\nu})^2$ and $(\na_\mu B^{\mu\nu})^2$ and
showing the reason to consider other terms in the action, as they
are required to provide consistency at the quantum level.

\section*{Note added}

In the recent paper \cite{Thierry-Mieg2023} there is a convincing
discussion of the global conformal symmetry in the flat-space limit. 
There is a possibility  \cite{Avdeev1993} that this symmetry is 
capable to explain the renormalizability of the model 
(\ref{actionflat}) without invoking local conformal symmetry.


\section*{Appendix A. \ Extended algebra of $\ga$-matrices}
\label{AppA}

Our notations include
$\ga^\mu = e^\mu_{\,a}\ga^a$, \ $\vp^{\mu\nu\al\be}
= \ep^{abcd} e^\mu_{\,a}e^\nu_{\,b}e^\al_{\,c}e^\be_{\,d}$ \
and
\beq
\ga^5 = - i \ga^0\ga^1 \ga^2 \ga^3
\,=\,\frac{i}{24}\,\vp_{\mu\nu\al\be}\ga^\mu \ga^\nu \ga^\al \ga^\be.
\label{gam1}
\eeq
All Greek indices are raised and lowered with the metric
$g^{\mu\nu}$ and its inverse $g_{\mu\nu}$.
Furthermore, we denote the antisymmetric combination as
\beq
\Si^{\mu\nu} \,=\,i\,\ga^{[\mu} \ga^{\nu]}
\,=\, \frac{i}{2}\,
\big( \ga^\mu \ga^\nu  - \ga^\nu \ga^\mu \big).
\label{gam2}
\eeq
The Clifford algebra for the curved-space gamma-matrices
has the form
\beq
\ga^\mu \ga^\nu  +  \ga^\nu \ga^\mu
\,=\,2 g^{\mu\nu}.
\label{gam3}
\eeq
The full basis in the space of spinor matrices is formed by
$I$, $\ga^\mu$, $\ga^5$, $\ga^5\ga^\mu$, and $\Si^{\mu\nu}$.
Using covariance and parity arguments, we can write the
general relation
\beq
\ga^\al \ga^\mu \ga^\nu \,=\,x_1  \ga^\al g^{\mu\nu}
+ x_2  \ga^\mu g^{\al\nu} + x_3  \ga^\nu g^{\al\mu}
+ i x_4\ga^5\vp^{\mu\nu\al\be}\, \ga_\be,
\label{gam4}
\eeq
where $x_{1,2,3,4}$ are unknown coefficients, which can be
easily found by contracting (\ref{gam4}) with $g_{\mu\al}$
and with $\vp_{\mu\nu\al\la}$. As a result, we arrive at the
well-known relation
\beq
\ga^\al \ga^\mu \ga^\nu \,=\, \ga^\al g^{\mu\nu}
- \ga^\mu g^{\al\nu} +  \ga^\nu g^{\al\mu}
+ i\ga^5\vp^{\mu\nu\al\be}\, \ga_\be,
\label{gam5}
\eeq
The consequent formulas are
\beq
&&
\ga^\al (\ga^\mu \ga^\nu - \ga^\nu \ga^\mu )
\,=\, 2 (g^{\mu\al} \ga^\nu  - g^{\nu\al} \ga^\mu)
+ 2i\ga^5\vp^{\mu\nu\al\be}\, \ga_\be,
\nn
\\
&&
(\ga^\mu \ga^\nu - \ga^\nu \ga^\mu ) \ga^\al
\,=\, 2 (g^{\nu\al} \ga^\mu  - g^{\mu\al} \ga^\nu )
+ 2i\ga^5\vp^{\mu\nu\al\be}\, \ga_\be,\nn
\\
&&
\ga^\al (\ga^\mu \ga^\nu - \ga^\nu \ga^\mu )
\,+\,
(\ga^\mu \ga^\nu - \ga^\nu \ga^\mu ) \ga^\al
\,=\, 4i\ga^5\vp^{\mu\nu\al\be}\, \ga_\be.
\label{gam6}
\eeq

Using the same approach we used in (\ref{gam4}), one can
derive the simplified form of the product of four (and more,
if necessary) gamma-matrices. For our purposes it is sufficient
to restrict the consideration by the antisymmetric version of the
product,
\beq
&&
\ga^{[\mu} \ga^{\nu]}\,\ga^{[\al} \ga^{\be]}
\,=\, i\ga^5 \vp^{\mu\nu\al\be}
\,-\,  \big( g^{\mu\al} g^{\nu\be} - g^{\nu\al} g^{\mu\be}  \big)
\nn
\\
&&
\qquad \qquad
+ \,\,i \big( \Si^{\mu\al} g^{\nu\be}
- \Si^{\nu\al} g^{\mu\be}
+ \Si^{\nu\be} g^{\mu\al}
-  \Si^{\mu\be} g^{\nu\al}\big).
\label{gam7}
\eeq

\section*{Appendix B. \
Some algebraic formulas for $\tilde{B}_{\mu\nu}$}
\label{AppB}

To elaborate necessary formulas involving $\tilde{B}_{\mu\nu}$,
in most of the cases one can use the contractions of antisymmetric
tensors, e.g.,
$\vp^{\mu\nu\al\be}\vp_{\mu\nu\rho\si}
= -2 \big(\de^\al_{\,\rho}\de^\be_{\,\si}
- \de^\al_{\,\si}\de^\be_{\,\rho}\big)$ and same method as in
Eq.~(\ref{gam4}), i.e., using symmetries, introducing free
coefficients and deriving them using contractions. Skipping the
details, let us present just the results, starting from the formula
which should be derived directly\footnote{In the
first version of this preprint there was a mistake in this formula,
which caused three wrong coefficients in the result for divergences
and required preparing an erratum to the published version.}
\beq
&&
\tilde{B}_{\mu\nu}\tilde{B}^{\al\be}
\,=\,
- \,B_{\mu\nu}B^{\al\be}
\,-\,
\frac12\,B_{\rho\si}^2
\big(\de_\mu^\al \, \de_\nu^\be
- \de_\nu^\al \, \de_\mu^\be\big)
\nn
\\
&&
\qquad \qquad
+\,\,\,
\de_\mu^\al B_{\nu\la} B^{\be\la}
- \de_\nu^\al B_{\mu\la} B^{\be\la}
+ \de_\nu^\be B_{\mu\la} B^{\al\la}
- \de_\mu^\be B_{\nu\la} B^{\al\la}\,.
\label{til1}
\eeq
A partial contraction gives
\beq
\tilde{B}^{\mu\nu}\tilde{B}^{\al\be} g_{\nu\be}
\,=\,
B^{\mu\nu}B^{\al\be} g_{\nu\be}
- \frac12\, B^{\rho\si}B_{\rho\si} g^{\mu\al}.
\label{til2}
\eeq
The consequences include the formulas
\beq
&&
\tilde{B}^{\mu\nu}\tilde{B}_{\mu\nu}
= - B^{\mu\nu}B_{\mu\nu},
\nn
\\
&&
C_{\al\be\mu\nu} \tilde{B}^{\al\be}\tilde{B}^{\mu\nu}
\,=\,
-\, C_{\al\be\mu\nu} B^{\al\be} B^{\mu\nu}
\label{til3}
\eeq
and, using also the basic reduction relations (\ref{K31red}),
\beq
&&
R_{\al\be\mu\nu} \tilde{B}^{\al\be}\tilde{B}^{\mu\nu}
\,=\,
2 R_{\al\be\mu\nu} \tilde{B}^{\al\mu}\tilde{B}^{\be\nu}
\nn
\\
&&
\qquad
\quad
=\,
\,-\, C_{\al\be\mu\nu} B^{\mu\nu}B^{\al\be}
\,+\, 2 B^{\mu\nu}B^{\al\be} R_{\mu\al} g_{\nu\be}
\, -\, \frac23\,R B^2_{\mu\nu}\, .
\label{til4}
\eeq

For the terms with covariant derivatives, after some
algebra, we get
\beq
&&
(\na_\al \tilde{B}_{\mu\nu}) (\na^\al \tilde{B}^{\mu\nu})
\,=\,
- \, (\na_\al B_{\mu\nu})^2,
\nn
\\
&&
(\na_\mu \tilde{B}^{\mu\nu}) (\na^\al \tilde{B}_{\al\nu})
\,=\,
(\na_\mu B^{\mu\nu})^2
- \frac12 (\na_\al B_{\mu\nu})^2
- \frac16\,R B^2_{\mu\nu} 
+ \frac12\, C_{\al\be\mu\nu} B^{\mu\nu}B^{\al\be},
\nn
\\
&&
(\na_\al \tilde{B}_{\mu\nu}) (\na^\mu \tilde{B}^{\al\nu})
\,=\,
-\,\frac12 (\na_\al B_{\mu\nu})^2 
+ (\na_\mu \tilde{B}^{\mu\nu})^2.
\label{til5}
\eeq

Finally, for the quartic terms, the reduction formulas are
$(\tilde{B}_{\mu\nu} \tilde{B}^{\mu\nu})^2
=(B_{\mu\nu} B^{\mu\nu})^2$ and
\beq
&&
\tilde{B}_{\mu\nu} B^{\mu\nu}
\tilde{B}_{\al\be} B^{\al\be}
\,=\,
-\,2\,\big(B_{\mu\nu} B^{\mu\nu}\big)^2
\,+\,4\,B_{\mu\nu} B^{\nu\al}
B_{\al\be} B^{\be\mu},
\nn
\\
&&
\tilde{B}_{\mu\nu} \tilde{B}^{\nu\al}
\tilde{B}_{\al\be} \tilde{B}^{\be\mu}
\,=\,
B_{\mu\nu} B^{\nu\al}
B_{\al\be} B^{\be\mu}.
\label{til7}
\eeq
The last relation can be derived either using (\ref{til1}) or
directly, by using the inverse to (\ref{dualB}),
\beq
\frac12\,\vp_{\mu\nu\al\be} \,\tilde{B}^{\al\be}
\,=\,B_{\mu\nu}\,.
\label{til8}
\eeq

\section*{Acknowledgements}

I am grateful to I.L. Buchbinder for useful correspondence and
to V.F. Barra for his participation at the early stage of the work.
The letters received after the first version of the preprint with
this work were very useful and greatly appreciated. The author
acknowledges important partial support from Conselho Nacional
de Desenvolvimento Cient\'{i}fico e Tecnol\'{o}gico - CNPq
under the grant 303635/2018-5.



\begin{thebibliography}{m}

\renewcommand{\baselinestretch}{1.0}
\small

\bibitem{OgiPolu67} V.I. Ogievetsky and I.V.~Polubarinov,
\textit{The notoph and its possible interactions,}
Yad. Fiz. \textbf{4} (1966) 216. 
[Sov. J. Nucl. Phys. \textbf{4} (1967) 156].

\bibitem{KalbRamon} M. Kalb and P. Ramond,
\textit{Classical direct interstring action,}
Phys. Rev. \textbf{D9} (1974) 2273.

 \bibitem{Damour1992} T.~Damour, S.~Deser and J.~G.~McCarthy,
\textit{Nonsymmetric gravity theories: Inconsistencies and a cure,}
Phys. Rev. \textbf{D47} (1993) 1541, 
gr-qc/9207003.

\bibitem{Avdeev1993} L.V.~Avdeev and M.V.~Chizhov,
\textit{Antisymmetric tensor matter fields: An Abelian model,}
Phys. Lett.  \textbf{B321} (1994) 212, 
hep-th/9312062.

\bibitem{Avdeev1994}
L.V.~Avdeev and M.V.~Chizhov,
\textit{A Queer reduction of degrees of freedom,}
Phys. Part. Nucl. Lett. \textbf{2} (2005) 7, 
hep-th/9407067.

\bibitem{Chizhov2017}
D.~P.~Kirilova and V.~M.~Chizhov,
\textit{Chiral tensor particles in the early Universe
\textemdash{} Present status,}
Mod. Phys. Lett.  \textbf{A32} (2017) 1750187,
arXiv:1711.07895.

\bibitem{GitmanSaa}
E.S.~Fradkin and D.M.~Gitman,
\textit{Path integral representation for the relativistic particle
propagators and BFV quantization,}
Phys. Rev. \textbf{D44} (1991) 3230;   
\\
D.M.~Gitman and A.~V.~Saa,
\textit{Quantization of spinning particle with anomalous
magnetic momentum,}
Class. Quant. Grav. \textbf{10} (1993) 1447, 
hep-th/9209086;
\textit{Pseudoclassical model of spinning particle with
anomalous magnetic momentum,}
Mod. Phys. Lett.  \textbf{A8} (1993) 463. 
hep-th/9208049.

\bibitem{Pasti1995} P.~Pasti, D.P.~Sorokin and M.~Tonin,
\textit{Space-time symmetries in duality symmetric models,}
Contribution to: Workshop on Gauge Theories, Applied Supersymmetry,
and Quantum Gravity, hep-th/9509052.

\bibitem{Quevedo1996} F.~Quevedo and C.A.~Trugenberger,
\textit{Phases of antisymmetric tensor field theories,}
Nucl. Phys.  \textbf{B501} (1997) 143, 
hep-th/9604196.

\bibitem{Buch2008} I.L.~Buchbinder, E.N.~Kirillova and N.G.~Pletnev,
\textit{Quantum equivalence of massive antisymmetric tensor field
models in curved space,}
Phys. Rev.  \textbf{D78} (2008) 084024,
arXiv:0806.3505.

\bibitem{Siegel1999} W.~Siegel,
\textit{Fields,}
(On-line advanced textbook) hep-th/9912205.

\bibitem{Altschul2009}
B.~Altschul, Q.G.~Bailey and V.A.~Kostelecky,
\textit{Lorentz violation with an antisymmetric tensor,}
Phys. Rev. \textbf{D81} (2010) 065028,
arXiv:0912.4852.

\bibitem{Albert}
J.F.~Assun\c{c}\~ao, T.~Mariz, J.R.~Nascimento and A.Y.~Petrov,
\textit{Dynamical Lorentz symmetry breaking in a tensor
bumblebee model,}
Phys. Rev. \textbf{D100} (2019)  085009,
arXiv:1902.10592.

\bibitem{Aashish2018}
S.~Aashish, A.~Padhy, S.~Panda and A.~Rana,
\textit{Inflation with an antisymmetric tensor field,}
Eur. Phys. J. \textbf{C78} (2018) 
887;
arXiv:1808.04315. 

\bibitem{Panda2023}
A.~Ajitha and S.~Panda, 
\textit{Inflation using a triplet of antisymmetric tensor fields,}
Eur. Phys. J. \textbf{C83} (2023) 770,
arXiv:2212.13508.

\bibitem{Sezgin1980}
E.~Sezgin and P.~van Nieuwenhuizen,
\textit{Renormalizability properties of antisymmetric tensor
fields coupled to gravity,}
Phys. Rev. \textbf{D22} (1980) 301.

\bibitem{Duff1980}
M.~J.~Duff and P.~van Nieuwenhuizen,
\textit{Quantum inequivalence of different field representations,}
Phys. Lett. \textbf{B94} (1980) 179. 

\bibitem{Tiberio-antis}
T.~de Paula Netto and I.L.~Shapiro,
\textit{Non-local form factors for curved-space antisymmetric
fields,}
Phys. Rev. D \textbf{94} (2016) 024040,
arXiv:1605.06600.

\bibitem{Aashish2018-int} S.~Aashish and S.~Panda,
\textit{Covariant effective action for an antisymmetric tensor field,}
Phys. Rev. \textbf{D97} (2018) 125005,
arXiv:1803.10157.

\bibitem{FrTs-superconf} E.S. Fradkin, and A.A. Tseytlin,
{\it One-loop beta function in conformal supergravities,}
Nucl. Phys. {\bf B203} (1982) 157. 

\bibitem{Fradkin1985}
E.S.~Fradkin and A.A.~Tseytlin,
\textit{Conformal supergravity,}
Phys. Rept. \textbf{119} (1985) 233. 

\bibitem{Branson} T.P.~Branson,
\textit{Conformally covariant equations on differential forms,}
Comm. Part. Diff. Equations \textbf{7} (1982) 393; 
\textit{Differential operators canonically associated to a
conformal structure,}
Math. Scandinavica \textbf{57} (1985) 293.

\bibitem{Erdmenger1997} J.~Erdmenger,
\textit{Conformally covariant differential operators:
Properties and applications,}
Class. Quant. Grav. \textbf{14} (1997) 2061, 
hep-th/9704108.

\bibitem{Barbashov1983}
B.M.~Barbashov and A.A.~Leonovich,
\textit{Conformally invariant theory of the vector and
antisymmetric tensor fields,}
Preprint JINR-P2-83-524 (1983).

\bibitem{Stud}
D.F.~Carneiro, E.A.~Freiras, B.~Goncalves, A.G.~de Lima
and I.L.~Shapiro,
\textit{On useful conformal tranformations in general relativity,}
Grav. Cosmol. \textbf{10} (2004) 305, 
gr-qc/0412113.

\bibitem{Penrose64} R.~Penrose,
\textit{Conformal treatment of infinity,}
(Les Houches Summer School of Theoretical
Physics, Editors B. De Witt and C. DeWitt),
reprinted in Gen. Rel. Grav. \textbf{43} (2011) 901.

\bibitem{Chernikov68}
N.A.~Chernikov and E.A.~Tagirov,
\textit{Quantum theory of scalar fields in de Sitter space-time,}
Ann. Inst. H. Poincare Phys. Theor. \textbf{A9} (1968) 109.

\bibitem{Paneitz} S. Paneitz,
{\it A quartic conformally covariant differential
operator for arbitrary pseudo-Riemannian manifolds,}
MIT preprint, 1983;
SIGMA {\bf 4} (2008) 036,
arXiv:0803.4331.

\bibitem{Paci2023} G.~Paci, D.~Sauro and O.~Zanusso,
\textit{Conformally covariant operators of mixed-symmetry
tensors and MAGs,}
Class. Quant. Grav. \textbf{40} (2023) 215005,

\bibitem{Hamada} K.j.~Hamada,
\textit{Integrability and scheme independence of
even-dimensional quantum geometry effective action,}
Prog. Theor. Phys. {\bf 105} (2001) 673, hep-th/0012053.

\bibitem{Int-6d} F.M. Ferreira and I.L. Shapiro,
{\it Integration of trace anomaly in $6D$,}
Phys. Lett. {\bf B772} (2017) 174, 
arXiv:1702.06892.

 \bibitem{Schwarz1978} A.S.~Schwarz,
\textit{The partition function of degenerate quadratic functional
and Ray-Singer invariants,}
Lett. Math. Phys. \textbf{2} (1978) 247, 
\textit{The partition function of a degenerate functional,}
Commun. Math. Phys. \textbf{67} (1979) 1, 

\bibitem{Grisaru1984}
M.T.~Grisaru, N.K.~Nielsen, W.~Siegel and D.~Zanon,
\textit{Energy Momentum Tensors, Supercurrents, (Super)traces
and Quantum Equivalence,}
Nucl. Phys. \textbf{B247} (1984) 157.

\bibitem{Buchbinder1988}
I.L.~Buchbinder and S.M.~Kuzenko,
\textit{Quantization of the classically equivalent theories
in the superspace of simple supergravity and quantum equivalence,}
Nucl. Phys. \textbf{B308} (1988) 162. 

\bibitem{Duff-94} M.J. Duff,
{\it Twenty years of the Weyl anomaly,}
Class. Quant. Grav. {\bf 11} (1994) 1387, hep-th/9308075.

\bibitem{PoImpo} I.L.~Shapiro,
\textit{Effective action of vacuum: semiclassical approach,}
Class. Quant. Grav. {\bf  25} (2008) 103001, arXiv:0801.0216.

\bibitem{OUP} I.L.~Buchbinder and I.L.~Shapiro,
\textit{ Introduction to quantum field theory with applications to
quantum gravity,} (Oxford University Press, 2021).

\bibitem{tmf} I.L. Buchbinder,
{\it On renormalization group equations in curved space-time,}
Theor. Math. Phys. {\bf 61} (1984) 393.

\bibitem{DeWitt65} B.S. DeWitt,
{\it Dynamical theory of groups and fields,}
(Gordon and Breach, 1965).

\bibitem{birdav} N.D.~Birell and P.C.W.~Davies,
{\it Quantum fields in curved space},
(Cambridge Univ. Press, Cambridge, 1982).

\bibitem{Cheng1973}
T.P.~Cheng, E.~Eichten and L.~F.~Li,
\textit{Higgs phenomena in asymptotically free gauge theories,}
Phys. Rev. \textbf{D9} (1974) 2259.

\bibitem{rie} R.J. Riegert,
{\it A nonlocal action for the trace anomaly,}
Phys. Lett. {\bf 134B} (1984) 56.

\bibitem{FrTs84} E.S. Fradkin and A.A. Tseytlin,
{\it Conformal anomaly in Weyl theory and anomaly free
superconformal theories,}
Phys. Lett. {\bf 134B} (1984) 187.

\bibitem{Mottola-2017} E.~Mottola,
\textit{Scalar gravitational waves in the effective theory of gravity,}
JHEP \textbf{07} (2017) 043,
[erratum: JHEP \textbf{09} (2017) 107]
arXiv:1606.09220.

\bibitem{anomaly-2004}
M. Asorey, G. de Berredo-Peixoto and I.L. Shapiro,
\textit{Renormalization ambiguities and conformal anomaly
in metric-scalar backgrounds,}
Phys. Rev. {\bf D74} (2006) 124011, hep-th/0609138;
\\
M.~Asorey, W.C.~Silva, I.L.~Shapiro and P.R.B.~d.~Vale,
\textit{Trace anomaly and induced action for a metric-scalar
background,}
Eur. Phys. J.  \textbf{C83} (2023) 157,
arXiv:2202.00154.

\bibitem{AtA} G.H.S.~Camargo and I.L.~Shapiro,
\textit{Anomaly-induced vacuum effective action with torsion:
Covariant solution and ambiguities,}
Phys. Rev. \textbf{D106} (2022) 045004,
arXiv:2206.02839.

\bibitem{Thierry-Mieg2023} J.~Thierry-Mieg and P.D.~Jarvis,
\textit{Conformal invariance of antisymmetric tensor field theories
in any even dimension,}
J. Phys.  \textbf{A58} (2025) 055402,
arXiv:2311.01701.

\end{thebibliography}
\end{document}